\documentclass[submission,copyright]{eptcs}
\providecommand{\event}{Graphs as models 2017} 
\providecommand{\dragan}{Bo{\v{s}}na{\v{c}}ki }
\newcommand{\circleT}[1]{\raisebox{.5pt}{\textcircled{\raisebox{-.9pt} {#1}}}}
\usepackage{breakurl}             
\usepackage{underscore}           
\usepackage{color, soul}
\usepackage{graphicx}
\usepackage{float}
\usepackage{epstopdf}
\usepackage{xspace}
\usepackage{wrapfig}
\usepackage{tikz}
\usepackage{verbatim}
\usepackage{pgfplots}

\newcommand{\GPUexplore}{\textsc{GPUexplore}\xspace}

\title{Analysing the Performance of GPU Hash Tables for State Space Exploration}
\author{Nathan Cassee
\institute{Eindhoven University of Technology\\ Eindhoven, The Netherlands}
\email{N.W.Cassee@student.tue.nl}
\and
 \quad\qquad Anton Wijs
 \institute{\quad \qquad Eindhoven University of Technology\\
\quad \qquad Eindhoven, The Netherlands}
\email{ \quad\qquad A.J.Wijs@tue.nl}
}
\def\titlerunning{Analysing the Performance of GPU Hash Tables for State Space Exploration}
\def\authorrunning{N.W. Cassee \& A.J. Wijs}
\begin{document}
\maketitle

\begin{abstract}
\textbf{Abstract:} In the past few years, General Purpose Graphics Processors (GPUs) have been used to significantly speed up numerous applications. One of the areas in which GPUs have recently led to a significant speed-up is model checking. In model checking, state spaces, i.e., large directed graphs, are explored to verify whether models satisfy desirable properties. \GPUexplore is a GPU-based model checker that uses a hash table to efficiently keep track of already explored states. As a large number of states is discovered and stored during such an exploration, the hash table should be able to quickly handle many inserts and queries concurrently. In this paper, we experimentally compare two different hash tables optimised for the GPU, one being the \GPUexplore hash table, and the other using Cuckoo hashing. We compare the performance of both hash tables using random and non-random data obtained from model checking experiments, to analyse the applicability of the two hash tables for state space exploration. We conclude that Cuckoo hashing is three times faster than \GPUexplore hashing for random data, and that Cuckoo hashing is five to nine times faster for non-random data. This suggests great potential to further speed up \GPUexplore in the near future.
\end{abstract}

\section{Introduction}

General Purpose Graphics Processors (GPUs) have been used to significantly speed up computations in numerous application domains. Contrary to CPUs, GPUs can handle many thousands of parallel threads, allowing for a great increase in parallel processing of data. This increase has opened up new possibilities to improve the performance of algorithms, as significant speedups can be achieved by using optimised algorithms and lock-free data structures. \par 
One of the domains to which GPU computing has been applied in the last few years is model checking. Model checking involves taking a model of an (often concurrent) system and verifying whether certain properties are satisfied by that model. The semantics of such models can be expressed in a \emph{state space}, which is a directed graph (often sparse), with labels either on the edges or the nodes. During model checking, such a state space is constructed by interpreting the model. Starting from the initial state of the model, exploration continues until either all reachable states have been explored, or a counter-example to the property being verified has been encountered~\cite{principlesofmodelchecking}. This operation can be very resource-demanding, as state spaces
tend to grow exponentially as the number of concurrent components in a model grows linearly. This problem is
commonly referred to as the \emph{state space explosion problem}. For this reason, the use of modern parallel architectures is very appealing, since it could make the analysis of large state spaces more practically feasible.
 
In recent years, approaches have been developed to perform state space exploration either using a combination of the CPU and the GPU, or specifically using the GPU~\cite{Spin, Edelkamp, bosnacki.edelkamp.2010, wijs.gpubisim, wijs.cav, WijKaBo16,GPUExplore, GPUExplore2, singaporePaper}. Most approaches use an exploration strategy similar to Breadth-First Search (BFS).


A data structure commonly used during state space exploration is the \emph{hash table} \cite{algorithms}. It is used to keep track of the states encountered so far. Accessing the hash table tends to be done very frequently during model checking, so in a parallel setting, it is critical that the hash table being used can very efficiently handle parallel accesses. For this reason, first of all, it needs to be \emph{lock-free}, and second of all, it needs to have a suitable hashing mechanism. Even though identifying which hash tables are most suitable for model checking is an important undertaking, for GPUs, this has not been thoroughly done in the past, or at least, no reports have been published on it. In this paper, we discuss such a thorough comparison.

Only a very few different hash tables have been proposed so far for GPUs. A notable one is the one by Alcantara \textit{et al.}~\cite{AlcChpt}. They propose to use a version of \emph{Cuckoo hashing}~\cite{cuckoo} optimised for parallel execution using 64-bit data elements. It has been adopted in the standard CUDPP library\footnote{CUDPP library: \url{http://cudpp.github.io.}}. In addition, Wijs and \dragan developed their own custom hash table for their parallel state space exploration tool \GPUexplore~\cite{GPUExplore,GPUExplore2}.

Alcantara \textit{et al.}\ compared the performance of their hash table to the building and querying of a sorted array. The keys and values used by Alcantara \textit{et al.}\ to test the performance of Cuckoo hashing are randomly generated integers \cite{phdAlc}. Consequently, the performance of their Cuckoo hash table has previously only been analysed using sequences of random and unique integers, and not using non-random sequences stemming from real data. Since the performance of a hash table can be influenced by different patterns encountered in the input data, for instance, the frequency at which the same values are encountered again and again, it is useful to consider non-random data as well.

Beides this, for the hash table implemented in \GPUexplore, no isolated performance evaluation has been done in the past. Instead of Cuckoo hashing, the hash table implemented for \GPUexplore uses another way to resolve collisions and instead of operating on 64-bit elements, it supports elements of arbitrary length.

This paper addresses reports the results of a performance evaluation between Cuckoo hashing and \GPUexplore hashing. We analyse Cuckoo hashing and \GPUexplore hashing with both random and non-random datasets. The non-random data originates from actual state space explorations performed by a model checker, in which the revisiting of states has been recorded. This data allows us to compare the applicability of the two hash tables for state space exploration. \par 

The structure of the paper is as follows. Related work is discussed in Section 2. In Section 3, first a brief explanation of GPUs is given, which is particularly focussed on NVIDIA GPUs and the CUDA programming interface, as we employed those GPUs for our experiments. Furthermore, Section 3 also provides an overview of the approach used in \GPUexplore to do model checking. A detailed description of the two hash tables is given in Section 4. The experimental setup is explained in Section 5. In Section 6, the results of the comparisons are discussed, and finally, conclusion and pointers to future work are given in Section 7. 

\section{Related work}

There are several papers introducing lock-free hash tables for GPUs~\cite{AlcChpt,PerfEvlLockFree,linkedList, pres}. However, in those papers only limited performance evaluations have been conducted. Specifically,  the evaluations that have been performed involved sequences of equally distributed random values. This is the case for the performance evaluations conducted in the dissertation of Alcantara \cite{phdAlc}, and the performance evaluation conducted by Misra and Chaudhur \cite{PerfEvlLockFree}. \par 

The hash table described by Alcantara \textit{et al.}~\cite{phdAlc,AlcChpt} uses Cuckoo hashing and is implemented in the CUDA library CUDPP. In the work of Alcantara \textit{et al.}, different aspects of Cuckoo hashing are thoroughly analysed. Various aspects influencing the performance of the hash table are analysed and compared, including the size of the hash table and how long an insertion on the hash table tends to take.
However, no performance evaluation on non-random data for Cuckoo hashing on the GPU exists in the literature. \par 

Other lock-free hash tables are those proposed by Moazeni \& Sarrafzadeh~\cite{linkedList} and by Bordawekar \cite{pres}. Unfortunately, to the best of our knowledge, no implementations are publicly available.

Regarding GPU accelerated model checking, relevant work is the paper by Bartocci \textit{et al.}~\cite{Spin}, in which they introduce modifications of the SPIN model checker that take advantage of a GPU architecture. They experience significant performance increases for larger problems, for which concurrency of the GPU starts to pays off. The hash table they use is the one by Alcantara \textit{et al}., which means that they are restricted to analysing state spaces containing states that can be stored in 64 bits.

Edelkamp \textit{et al.}~\cite{Edelkamp} investigated the applicability of GPUs for probabilistic model checking, a process in which matrix-vector multiplications are performed and systems of linear equations are solved. Modifications to the algorithm such that these operations are performed on a GPU can cause speed-ups of 18 times compared to executions on the CPU.
 
Furthermore, Wu.\ \textit{et al.}\ investigated the use of GPUs for on-the-fly state space exploration using Cuckoo hashing~\cite{singaporePaper}, and Wijs and \dragan developed \GPUexplore, which uses a custom collision revolvement scheme to maintain which states have already been visited. The latter tool achieves a performance increase of, on average, 120 times compared to CPU based state space exploration, in cases where state spaces consisting of at least a million states are analysed~\cite{GPUExplore,GPUExplore2}.

Finally, Edelkamp \textit{et al.}\ investigated the applicability of GPUs to speedup the generation of very large state spaces using BFS~\cite{perfectGPUHashing}. In their work, Edelkamp \textit{et al.}\ utilise a perfect hashing function to keep track of the current depth of the BFS, so that during iterations over the state space, only the currently open states are considered.

\section{Background}
\subsection{CUDA}

In the 2000s, development of more programmable GPUs started. The change from fixed pipeline graphical units to programmable graphical processors opened up new possibilities for software developers. GPUs have significantly more processing units than CPUs, allowing more instructions to be executed in parallel.

The introduction of more programmable GPUs has made it possible to use GPUs for applications other than graphical calculations. Before the arrival of general purpose graphics processors, the only way to use GPUs for more general purpose applications was by transforming the problem to a problem of triangles and polygons, and then solving this problem using the available graphics Software Development Kits (SDK).

The increased capability of GPUs led Stanford researchers to define GPUs as stream processors. In turn, this has made it possible to compile C code to be executed on the GPU. NVIDIA expanded upon this concept and in 2006 released the first version of CUDA. CUDA is an SDK that allows code in a high-level language such as C or C++ to be compiled for, and executed on, an NVIDIA GPU. \par  

Unlike CPUs, GPUs are built to take large batches of data and execute the same, short, sequence of operations in parallel on all elements of that data. Because of this, GPUs often consist of several hundreds of processors, grouped in Streaming Multiprocessors (SMs), such that a large number of threads can concurrently execute instructions. Therefore, lock-free algorithms optimised for execution on a GPU often achieve significant speed-ups. Matrix addition is an example of an operation that can easily be executed in parallel, which each single thread computes the value of a single cell of the output matrix. Because of the large number of threads that can be executed in parallel, the resulting matrix can be computed much faster than when it has to be constructed sequentially.

A CUDA program often has the following structure: Preparatory code that can not be executed in parallel is sequentially executed on the host CPU. Such code often prepares input data for parallel processing. After processing the input data a \emph{kernel} is launched, which is a method written for the GPU. After the kernel has finished executing, the output of the kernel can be copied back to the host memory and reported to the end user.

Launching a kernel for execution on the GPU is done by specifying how many \emph{thread blocks} should execute the kernel and the amount of threads that each block should consist of. The threads in all blocks all execute the same kernel, and together, they form a \emph{grid}~\cite{cuda}.

Within a block, threads execute in \emph{warps} consisting of 32 threads. Within a warp, the threads execute in lockstep. This means that if the execution of a kernel involves an if statement, and, for instance, only 10 threads enter the body of the if-statement, then the other 22 threads are paused while the 10 active threads execute that body. More generally stated, kernels that cause the threads in a warp to branch incur performance penalties.
Finally, memory accesses are scheduled on the half-warp level, meaning that if either the first or last 16 threads in a warp wish to access a coalesced part of the memory, they will do so in lock-step.

\begin{wrapfigure}{r}{0.5\textwidth}
\centering
\includegraphics[scale=.3]{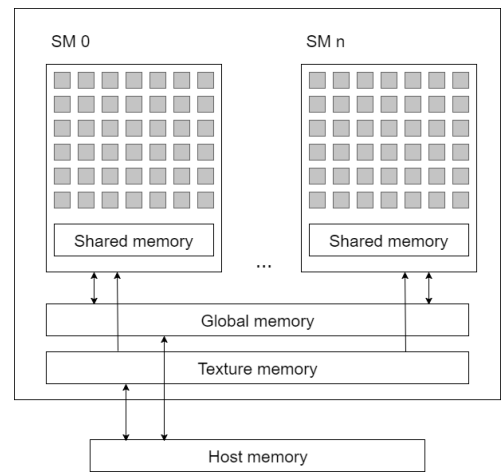}
\caption{CUDA memory model, the small gray blocks are individual threads}
\label{fig:memmodel}
\end{wrapfigure}

In addition to threading, the memory model of the GPU is different compared to the memory hierarchy of a CPU. Figure \ref{fig:memmodel} illustrates the CUDA memory model. The memory layout GPU consists out of several memory blocks. Most notably, texture memory and global memory are available to all SMs. While access to the texture memory is faster than access to global memory, write accesses to the former can only be performed from the CPU side. On the other hand, the global memory has a higher latency, but can both be read and written to from the SMs.


Next to the registers, which is fast on-chip local memory for each individual thread, every thread block can use shared memory, which is memory that is shared between all threads in a block. The advantage of shared memory is that it is also on-chip, and therefore closer to the SM than global memory, and therefore faster to access. However, the shared memory of a block is not accessible by threads outside of that block. This makes shared memory only useful for exchanging information between threads in the same block. \par 

In addition to having different forms of memory available to SMs, each type of memory also has different latency and throughput characteristics. Texture memory is faster to access than global memory. However, global memory is much larger than texture memory. For all types of device memory it holds that it is faster to perform coalesced memory access, e.g., a warp reading a continuous piece of global memory, instead of each thread accessing a distinctly different memory location. \par 

The division of threads into grids, blocks and warps and the unique memory model of CUDA has far reaching implications for algorithms that need to be optimised for execution on a GPU. However, the correct application of these concepts can lead to vast performance improvements.

\subsection{\GPUexplore}
In 2014, Wijs and \dragan developed a GPU powered on-the-fly model checking tool, called \GPUexplore~\cite{GPUExplore,GPUexplore-safety}, that initially on average achieved a 10 times speedup compared to CPU implementations. The latest version of \GPUexplore achieves an average speedup of 120 times for sufficiently large state spaces~\cite{GPUExplore2}. Graph traversal algorithms optimised for usage on a GPU already existed prior to the development of \GPUexplore, but these algorithms utilised the fact that the size of the final graph was already known before exploration started. However, for model checking, this is not a practical assumption, therefore \GPUexplore initially only knows the size of the model, which consists of a finite number of automata
in parallel composition.
Because of this, the parallelisations proposed in the literature for graph traversal could not directly be applied for on-the-fly state space exploration. \par 

As input, \GPUexplore excepts a model of a concurrent system, consisting of a finite number of finite-state \emph{processes}. It explores the state space implied by this model using a Breadth-First Search (BFS) based search strategy, starting from the initial state of the system. The initial system state is a vector consisting of all the initial states of the individual processes. By combining the enabled transitions in each individual process, successor state vectors can be constructed. To keep track of discovered and explored states \GPUexplore maintains a set of open states and a set of closed states, respectively. The set of open states contains the states for which the successors still need to be identified, and the set of closed states is the set of states for which this has already been done. In addition to a finite number of processes, an automaton representing a (negation of a) safety property can be provided. Recently, support for liveness properties has also been investigated~\cite{Wijs2016}, and support for partial order reduction to restrict exploration of state spaces has  been added~\cite{gpupor}. In the future, support for timed behaviour is planned~\cite{wijs:drt} and other forms of state space reduction~\cite{torabidashti.wijs:beamsearch}.\par 

The implementation of \GPUexplore solves several practical aspects related to state space exploration. For instance, the papers address how the input model is read and stored in memory, and how the concurrent processes in a model should be combined to discover all the reachable states of the complete system.  

As both the memory model and the threading model of CUDA drastically differ from CPUs there are several notable implementation details. Due to the fact that memory latency is high, a reduction of the number of required memory accesses can lead to significant performance increases. Firstly, the individual processes in the model are stored in texture memory, as this information never needs to be updated during exploration. Because of this accessing the states and outgoing transitions of a state of an individual process only incurs minimal memory latency. \par 
Secondly, as the number of threads available on a GPU is much higher than on a CPU, \GPUexplore gives each individual thread a very small amount of work to perform. When the successors of a system state need to be identified, a group of threads cooperate, and each individual thread fetches the relevant outgoing transitions of the current state of a particular process. Together, the threads in a group ensure that the correct result is computed. \par 

To implement the open and closed sets, \GPUexplore uses a hash table to store visited states.
Since the encoding of a state vector depends on the number of processes in the concurrent system model, and the number of local states each process can be in, a single 32 or 64-bit integer is often too small to contain a vector. Therefore, the hash table of \GPUexplore is able to correctly store elements consisting of multiple 32-bit integers, where each element is a state in the concurrent system. \par 
The hash table used by \GPUexplore to keep track of the visited and open sets is stored in global memory of the GPU. As memory latency for the global memory is high, especially for uncoalesced memory accesses  \cite{cuda}, the hash table provides the guarantee that an item can be located in a worst case constant number of steps. \par 
\GPUexplore does not use the hash table of Alcantara \textit{et al.}, as it only supports atomic insertions of elements consisting of single 32 or 64-bit integers. If Cuckoo hashing is used to insert longer elements, race conditions tend to frequently occur~\cite{GPUexplore-safety}. This may lead to erroneous state vectors being stored in the hash table that were not really visited during exploration. Therefore, Wijs and \dragan developed \GPUexplore hashing which is suited for vectors of arbitrary length.


\section{Hashtables}

In this section, the two hash tables that have been evaluated in this paper are discussed.

\subsection{Cuckoo hashing on the GPU}

In 2001, Pagh and Rodler proposed Cuckoo hashing \cite{cuckoo}, a collision resolving scheme for hash tables. Contrary to collision resolving schemes \cite{algorithms} such as linear probing, quadratic probing or chaining, locating a key in the hash table can be done in worst case constant time. The average performance of Cuckoo hashing is comparable to other hash table designs.

A Cuckoo hash table $\mathcal{H}_c$ has a constant number of hashfunctions $h_1$, .. $h_c$. Insertion of a key $k$ is initially done by hashing $k$ with $h_1$. For instance, let $h_1(k) = i$, then the key is inserted into $\mathcal{H}_c$ at position $i$. If there is no element at this position the insertion algorithm terminates. Otherwise, if there already is a key $k'$ at position $i$, $k$ is inserted into $\mathcal{H}_c$ at $i$ and $k'$ is \emph{evicted} and reinserted into the hash table at position $i'$, which is determined by the next hash function relevant for $k'$, that is, given that for some $a \in \{ 1, .., c\}$, we have $h_a(k') = i$, then $h_{a+1}(k')$ determines the new position $i'$. This process is repeated until an empty slot is encountered, a maximum number of so-called evictions is reached, or the number of hash functions was not sufficient to insert a particular key. In the latter two cases, the hash table is considered to be full.

\begin{figure}[t]
\centering
\includegraphics[scale=.45]{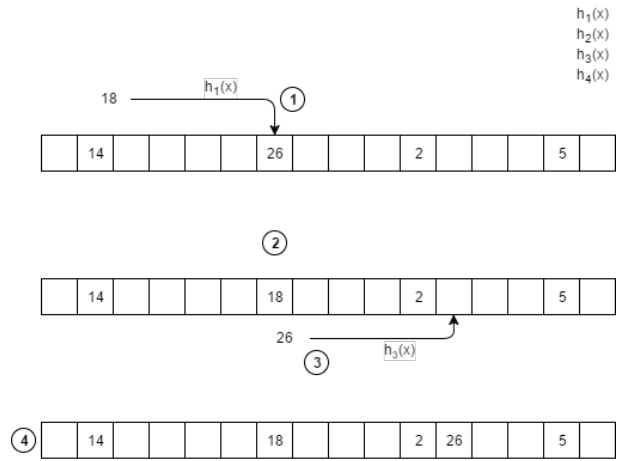}
\caption{Visual representation of an insertion into a hash table using Cuckoo hashing}
\label{fig:cuckoohash}
\end{figure}

As an example, the process of inserting key 18 into a Cuckoo hash table is illustrated in Figure~\ref{fig:cuckoohash}. As can be observed, in step \circleT{1}, the key is first hashed using the first hash function $h_1$ and inserted into the corresponding slot.
However, in this case there already is an element that has been inserted into this slot. That element is evicted from the table (step \circleT{2}) and should be re-inserted using its next hash function. Before re-inserting 26 the algorithm first determines which hash function was used to previously insert key 26 into the table. In this case 26 was inserted using hash function $h_2$, and therefore it is re-inserted using hash function $h_3$ (step \circleT{3}). Since there is no element in the slot that 26 is inserted into, the insertion terminates (step \circleT{4}).
In this case there was only one eviction, as rehashing 26 pointed to a free slot, however, would there have been another element at the final location of 26 that element would have been evicted as well. Such a sequence of evictions for a single insertion is called an \emph{eviction chain}.

To locate a key $k$ in the hash table, worst case, all possible slots indicated by hashing $k$ with $h_1$, .. $h_c$ need to be checked.
Hence, the query algorithm has a worst case constant running time, as at most $c$ steps have to be performed to locate a key in the hash table. Furthermore for insertions, Pagh and Rodler state that on average insertions perform similar to other hash tables using linear or quadratic probing.  \par 

Alcantara \textit{et al.}\ adapted the work of Pagh and Rodler to support the computational model of CUDA. Over the past few years Alcantara \textit{et al.}\ have proposed several variations on Cuckoo hashing, attempting to achieve  optimal performance \cite{phdAlc, AlcTwoLevelCuckoo, AlcChpt}. The overall best performing implementation largely resembles the originally proposed Cuckoo hash table by Pagh and Rodler. 
In that implementation, a kernel is launched for every batch of insert or query operations, where each element is processed by a single thread. This way the kernel makes optimal use of the available parallelism, allowing for a large amount of data to be processed concurrently.

Alcantara \textit{et al.}\ describe and compare different edge cases that can occur when using Cuckoo hashing. Most importantly, they discuss the size of the hash table compared to the size of the input data. Cuckoo hashing does not use any form of chaining, so as the hash table becomes more populated the length of eviction chains increases. Therefore, it is important to initially pick a hash table size that is sufficiently larger than the size of the input data. This ensures that eviction chains remain short even after almost all data has been inserted into the table.
Using experimental evaluation Alcantara \textit{et al.}\ determined $\mathcal{H}_{size} = 1.25n$ to be the ideal hash table size for input of size $n$. 
However, due to the nature of eviction chains and the constant number of hash functions it might always be the case that an element cannot be inserted as there might not be enough hash functions. In this case Alcantara \textit{et al.}\ propose to store such items in a small stash, and to attempt to re-insert elements from the stash if the stash becomes full. In case the stash is full and insertion of an element fails, the hash table has to be rebuilt, and new constants for the hash functions have to be generated. \par 
\begin{figure}[t]
\centering
\includegraphics[scale=.65]{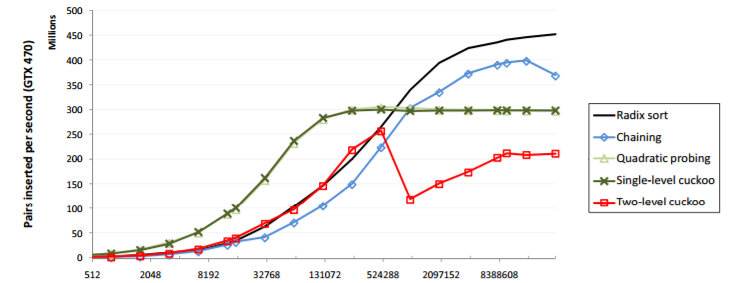}\\
\includegraphics[scale=.65]{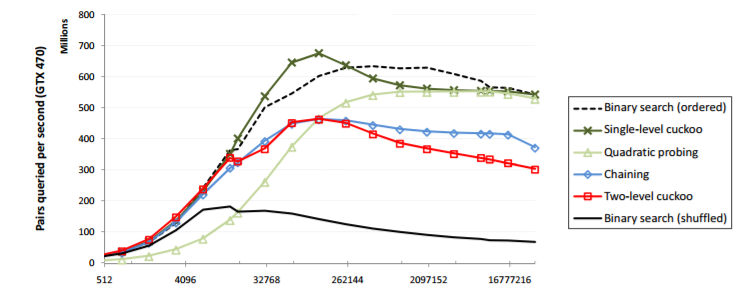}
\caption{Results of querying and inserting keys and values into several different hash tables~\cite{phdAlc}.}
\label{fig:alcresults}
\end{figure}

Figure~\ref{fig:alcresults} presents the original results as reported by Alcantara \textit{et al.}. They present two types of Cuckoo hash tables, one using a single level of Cuckoo hashing, the other using two, nested hash tables. As can be seen, using Cuckoo hashing on the GPU is faster when querying data and has a performance roughly equal to that of inserting when compared to other hash tables. Furthermore, when inserting data Alcantara \textit{et al.}\ found that for larger datasets, sorting the input sequence using radix-sort was faster that inserting all elements in a hash table, with Cuckoo hashing and quadratic probing being a close second \cite{phdAlc}.

\subsection{\GPUexplore implementation}

The hash table backing \GPUexplore resembles Cuckoo hashing in that it uses a constant number of hash functions. However, due to the information stored for each state that is encountered the \GPUexplore hash table needs to support storing vectors of 32-bit integers instead of single 32 bit elements.

Another difference is that instead of hashing to single slots, the \GPUexplore hash table hashes elements to \emph{buckets} of multiple slots. In this setup, each bucket can store several vectors. The size of a bucket has been set such that memory accesses by individual threads in a warp are grouped and optimised, as in GPU applications it is very important that memory accesses are coalesced as opposed to randomly scattered~\cite{cuda}. The size of a bucket for \GPUexplore hashing has been set to 32 integers, or equal to the warp-size. Furthermore, since atomic read and write operations are scheduled per half-warp, ensuring that buckets are aligned with warps helps to guarantee that vector reads and writes performed by one warp cannot be interrupted by another, which essentially boils down to those operations being atomic as well.

\begin{figure}[t]
\centering
\includegraphics[scale=.6]{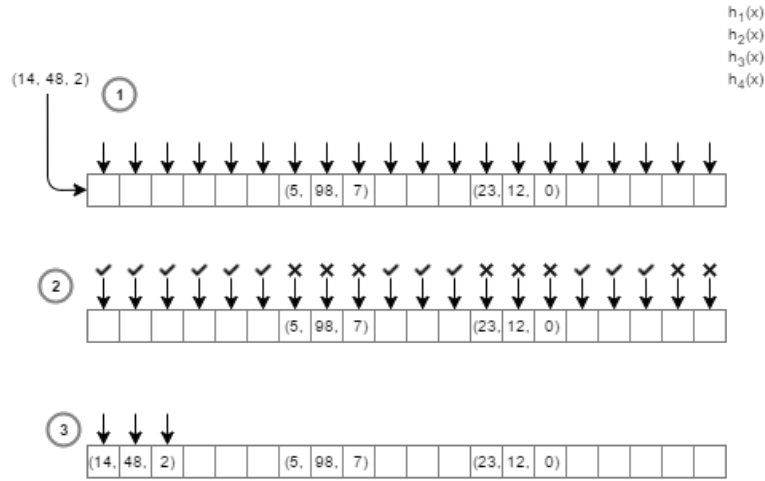}
\caption{Insertion of a vector into the bucket of a \GPUexplore hash table. }
\label{fig:gpuexplore}
\end{figure}

Figure \ref{fig:gpuexplore} illustrates an example of applying the insertion algorithm of \GPUexplore hashing to insert of vector consisting of three 32-bit integers. To insert a vector into the hash table it is first hashed by an entire warp of threads using the first hash function $h_1$. The corresponding bucket is then accessed by the entire warp, as can be seen in \circleT{1}, where each thread in the warp accesses a single slot in the bucket.
Then, every thread reports whether the slot that it is checking is free, or whether it has found an element of the vector that is being inserted \circleT{2}. Using warp primitives from CUDA these results are exchanged between threads such that every thread in the warp knows the result of all other threads in the warp. \par 
Using this information the threads can verify whether the vector that is being inserted is already present in the bucket. If this is the case all 32 threads can stop the insertion step as the element has already been inserted, otherwise a sufficiently large group of threads associated with single slots in the bucket attempts to atomically write the vector into the free sequence of slots (\circleT{3}). In this example, only three of the 32 threads are active while writing, and each thread in this case writes one 32-bit integer of the vector to a slot in the hash table. \par 
As the \GPUexplore hash table has been built with parallel insertions in mind it could be the case that between \circleT{2} and \circleT{3} another warp inserts a different vector into the targeted slot. In this case, the atomic insertion fails and the warp selects the next sequence of three free slots in the table, if present. The warp keeps doing this until all free slots have been tried, or until an atomic write completed successfully. \par 

There is no data available on how often such a conflict occurs, however, if this edge case is ignored state vectors are corrupted during the exploration phase of \GPUexplore and as a result an incorrect number of states is returned.

\par
If all free slots in a bucket have been tried unsuccessfully, or if the bucket had no free slots in the first place, the algorithm hashes the vector with the next hash function $h_2$ and probes the bucket returned by that function. The warp will keep doing this until either the vector has been placed, found or until there are no more hash functions that can be tried. In case the number of hash functions has been exhausted the algorithm reports that it considers the hash table to be full, as the vector could not be placed. \par 
Similar to Cuckoo hashing, attempting to find a vector in the \GPUexplore hash table has a worst-case constant time complexity. This is the case since at most $c$ buckets have to be probed to find the vector, and for each each bucket exactly 32 slots have to be checked. 

\section{Experimental setup}

To compare the performance of the \GPUexplore hash table and the Cuckoo hash table available in the CUDPP library, experiments have been performed on an Nvidia GPU. Both implementations have been tested using as input sequences of 32-bit integers. Two custom CUDA kernels have been written that execute the tests. Each kernel attempts to find and insert elements in one of the two hash tables, using the approporiate insertion operation. \par 

Both hash table implementations have several parameters that influence the running time of the insertion and querying algorithms. One of these parameters is the number of hash functions, which influences how many times an element can be rehashed before the hash table is considered full.
In addition, both hash tables also have a parameter that represents the size of the hash table versus the size of the dataset. For example, if there are ten unique elements, and a scale factor of $1.2$ defines that the hash table will have 12 slots.
As previously mentioned, for the Cuckoo hash table implementation of Alcantara \textit{et al.}, it has been suggested that about four hash functions and a hash table with size $1.25 * n$~\cite{AlcChpt} is ideal, where $n$ is the number of unique 32-bit integers in the input sequence. These same parameters have also been used for the benchmark experiments performed with the \GPUexplore hash table.

The tests were performed by first loading the sequence file into host memory, then the unique number of elements in the sequence is counted. Based on the unique number of elements in the sequence the hash table under test is initialized in the GPU memory with size 1.25 times the number of unique elements in the sequence file.
In addition, other information used by the hash table is copied to GPU memory, such as the hash constants used by the hash function. Finally, the sequence is copied to GPU memory and a kernel is started that attempts to insert all elements into the hash table under test.

For each element in the sequence it is first checked whether the element is already in the hash table, if the element is found in the hash table nothing is done. Otherwise, if the element does not exist in the hash table it is inserted using the insertion algorithm of the hash table under test. The time measured for each hash table is the time taken to either find or insert every single integer in the sequence, if an integer in the sequence is found it is not inserted in the table. Each experiment is conducted ten times, and the average of all ten runs is taken. \par 

While running the experiments we found out that enabling shared memory when launching the kernels negatively impacts performance, even if the shared memory is not actually used. Since shared memory is not required for the performance evaluation one set of experiments has been executed with shared memory disabled. However, as \GPUexplore requires shared memory during state space exploration  we have also analysed the performance of the two hash tables with shared memory enabled, so that the hash table best suited for \GPUexplore can be selected. 

\subsection{Modifications}

Both the Cuckoo hash table and the \GPUexplore hash table have been modified to be able to process the input sequence files. These modifications are explained in the following two subsections. 

\subsubsection{Cuckoo hashing}

In the original tests conducted by Alcantara \textit{et al.}, all values in the input sequences only appeared once. However, this is not realistic for state space exploration, as states tend to the encountered multiple times \cite{propertiesStateSpaces}.
However, the CUDPP implementation has been tailored for a priori known sequences of input values, and a precondition of the insertion method implemented in CUDPP is that no duplicate elements are ever inserted. For this reason, the insertion method proposed by Alcantara \textit{et al.}\ has been modified by us to also be able to handle parallel insertions of duplicate elements. This has been achieved by modifying the insertion method in such a way that it checks whether the element it is evicting is equal to the element it is inserting. If this is the case a parallel insertion is occurring, and the thread that has detected this parallel insertion terminates.  

\subsubsection{\GPUexplore} 

As the performance evaluations are executed on a list of 32-bit integers, the hash table of \GPUexplore has been modified to instead use single 32-bit integers as opposed to vectors of 32 bit integers. For the insertion method this means that a warp of 32 threads inserts a single 32-bit integer into a bucket. 

\subsection{Data}
Two different datasets have been used to compare the hash table implementations. One dataset is a sequence of randomly generated integers, while the second consists of integer sequences that represent the progress of actual sequential state space explorations, extracted from model checking executions. \par 

The random dataset has been generated using the random number generator in Python 3.6. Several different random datasets have been used to test the performance of the hash tables. Each randomly generated sequence has a length of 100 million integers; however, the range of values from which random numbers have been picked differs for each dataset. This variation influences how often particular values appear in the sequences. \par 

The non-random datasets have been obtained by running a modified version of the \textsc{Refiner}~\cite{refiner} model checking tool on several example models provided with \textsc{Refiner}. Three datasets have been generated using \textsc{Refiner}, the models used for these are HAVi-asyn, Sieve and ABP, which originate from the CADP toolbox distribution~\cite{Garavel2013}. The generated sequences vary in length from 35 million to 178 million integers. All three models can be downloaded together with \textsc{Refiner}.\footnote{\textsc{Refiner} is available at \url{http://www.win.tue.nl/~awijs}.} \par 
 
In total, 100 random integer sequences have been generated. During state space exploration, the same state is often encountered several times \cite{propertiesStateSpaces}, therefore, the generated sequences vary in how often every integer value occurs. \par

\section{Results} \label{sec:results}

The experiments have been performed on an Nvidia GT 750M GPU using the Kepler architecture with 2 streaming multiprocessors, 2 GBs of global memory and a total of 384 cores. The code has been compiled with CUDA 8 using compute capability 3.0 and the evaluations have been executed on Windows 10.  \par 

\begin{figure}[t]
	\centering
	\begin{tikzpicture}

	\begin{axis}[
	ylabel=Runtime (ms),
	xlabel=Duplication in sequence,
	tick label style={font=\footnotesize},
	xtick={0,100, ..., 1000},
	xmin=0, xmax=1000,
	ymin=0,
	ymax=8500,
	ytick={0,1000, ...,8000},
	tick align=outside,
	xtick pos=left,
	width=.6\textwidth,
	height=6cm,
	mark size=1.5pt,
	cycle list={%
		red,mark=*\\%
		blue,mark=square*\\%
		black,mark=triangle*\\%
		brown,mark=star\\%
		teal,mark=diamond*\\%
		orange,mark=*\\%
		violet,mark=square*\\%
		cyan,mark=triangle*\\%
		green!70!black,mark=start\\%
		magenta,mark=diamond*\\ %
	},
	]
	\pgfplotsinvokeforeach{gpuexploreshared, cuckooshared}{
		\addplot+ table[x=division,y=#1] {data/resultstikz.csv};
	}
	
	\legend{\GPUexplore shared, Cuckoo shared}
	\end{axis}
	
	\end{tikzpicture}

\vspace{.25cm}	

	\begin{tikzpicture}
	\begin{axis}[
	ylabel=Runtime (ms),
	xlabel=Duplication in sequence,
	tick label style={font=\footnotesize},
	xtick={0,100, ..., 1000},
	xmin=0, xmax=1000,
	ymin=0,
	ytick={0,1000, ...,8000},
	ymax=8500,
	tick align=outside,
	xtick pos=left,
	width=.6\textwidth,
	height=6cm,
	mark size=1.5pt,
	cycle list={%
		red,mark=*\\%
		blue,mark=square*\\%
		black,mark=triangle*\\%
		brown,mark=star\\%
		teal,mark=diamond*\\%
		orange,mark=*\\%
		violet,mark=square*\\%
		cyan,mark=triangle*\\%
		green!70!black,mark=start\\%
		magenta,mark=diamond*\\ %
	},
	]
	\pgfplotsinvokeforeach{gpuexplore, cuckoo}{
		\addplot+ table[x=division,y=#1] {data/resultstikz.csv};
	}
	
	\legend{\GPUexplore, Cuckoo}
	\end{axis}
	\end{tikzpicture}
	\caption{Results of evaluating the performance of Cuckoo hashing and \GPUexplore hashing on sequences of randomly generated numbers, with shared memory enabled (top) and disabled (bottom). Every sequence is 100,000,000 integers long, and the x-axis indicates how often a value occurs on average in the sequence.}
	\label{fig:randresults}
\end{figure}
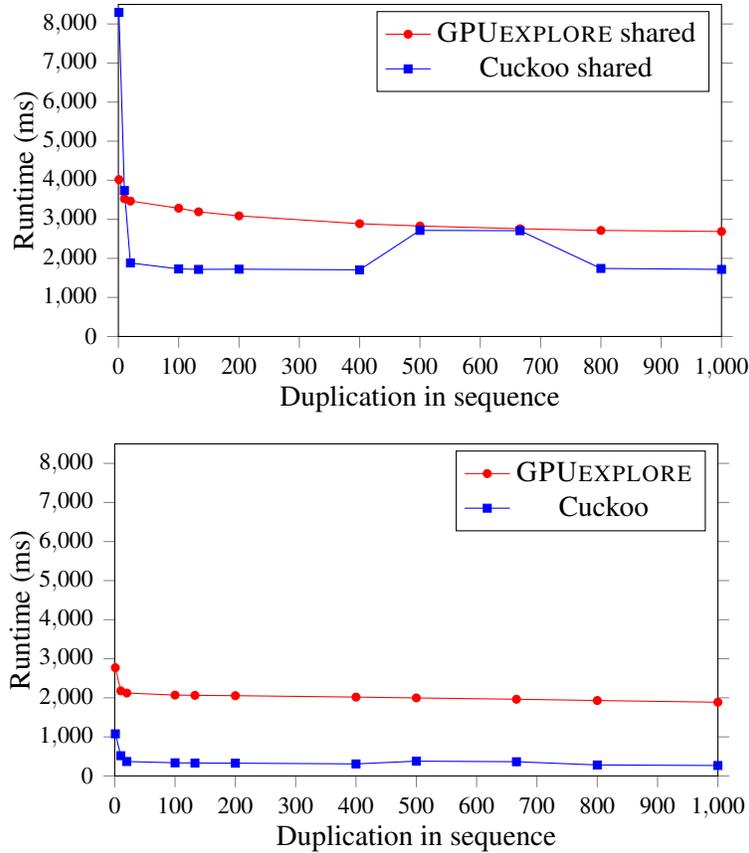

Datapoints have been recorded for a total of four different configurations. Both hash tables have been evaluated, and for each hash table tests have been run with shared memory enabled and disabled.  \par 

The results of analysing the performance of the hash tables on sequences of randomly generated data can be observed in Figure \ref{fig:randresults}. The y-axis displays the running time of each test in milliseconds and the x-axis displays how often an integer in the sequence occurs on average, where $n$ means that on average an integer occurred $n$ times in the entire sequence. Furthermore, the top figure shows the running time with shared memory enabled and the bottom figure shows the running time with shared memory disabled.\par 

Enabling shared memory appears to slow down the performance of the hash tables by roughly 1/3, with a peak slowdown where the shared memory version is 8 times slower than the non-shared memory version. With shared memory enabled fewer \emph{blocks} can be scheduled on a single SM, as each SM only has a limited amount of shared memory available. Therefore fewer threads operate simultaneously which means that fewer elements are inserted in parallel and therefore performance decreases.  \par 
For the tests executed with shared memory disabled it becomes clear that Cuckoo hashing is about three times faster than \GPUexplore hashing. Given that \GPUexplore hashing uses 32 threads to insert a single element, Cuckoo hashing uses a single thread for each integer this result is interesting, since the fact that \GPUexplore hashing uses 32 more threads to insert a single element than Cuckoo hashing implies that Cuckoo hashing could in theory be 32 times faster.

Besides this, it also becomes clear that insertions are more expensive than queries for both hash tables. As soon as duplicate elements are introduced in the sequence the running time to process the entire sequence decreases sharply. However, \GPUexplore hashing appears to execute a larger amount of constant work as the drop-off in running time is not as great as the drop-off for Cuckoo hashing.  \par 

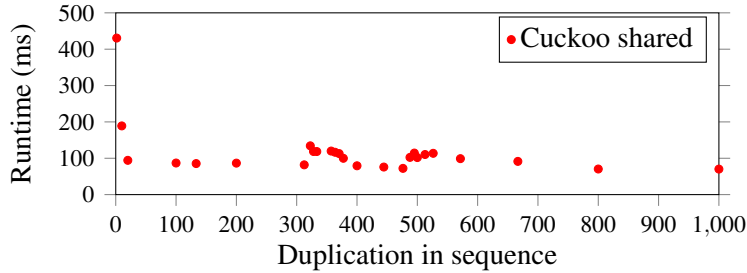
\begin{figure}[t]
\label{fig:resbump}
\centering
\begin{tikzpicture}
	\begin{axis}[
	ylabel=Runtime (ms),
	xlabel=Duplication in sequence,
	tick label style={font=\footnotesize},
	xtick={0,100, ..., 1000},
	xmin=-0, xmax=1000,
	ymin=0,
	ytick={0,100, ...,500},
	ymax=500,
	tick align=outside,
	xtick pos=left,
	width=.6\textwidth,
	height=4cm,
	mark size=1.5pt,
	cycle list={%
		red,mark=*\\%
		blue,mark=square*\\%
		black,mark=triangle*\\%
		brown,mark=star\\%
		teal,mark=diamond*\\%
		orange,mark=*\\%
		violet,mark=square*\\%
		cyan,mark=triangle*\\%
		green!70!black,mark=start\\%
		magenta,mark=diamond*\\ %
	},
	]
	\pgfplotsinvokeforeach{cuckooshared}{
		\addplot+[only marks] table[x=division,y=#1] {data/resbumptikz.csv};
	}
	
	\legend{Cuckoo shared}
	\end{axis}
	\end{tikzpicture}
\caption{Results of running the performance evaluation for Cuckoo hashing with shared memory enabled on a GTX 970. 
}
\end{figure}

The two bumps in the running time of Cuckoo hashing for n = 500 and n = 666.67 are as of yet unexplained. For some reason sequences inserted with Cuckoo hashing where each number occurs 450 to 700 times take on average 1.5 times longer to insert, compared to numbers occurring slightly less than 450 times and slightly more than 700 times. The experiments have been rerun several times using different sequences and on a different GPU, namely a GTX 970. Nevertheless, the bump in running time for Cuckoo hashing still manifests itself as can be seen in Figure \ref{fig:resbump}. However, in the non-random sequences, numbers do not tend to occur on average between 450 and 700 times. This suggests that for model checking problems, the observed performance bump may not heavily impact state space exploration. \par 

\begin{table}[th]
\centering
\caption{Performance evaluation between Cuckoo hashing and \GPUexplore hashing on non-random data}
\medskip
\label{tab:nonrandomres}
\begin{tabular}{|l|r|r|}
 \hline 
                                & Cuckoo hashing (ms) & \GPUexplore hashing (ms) \\ \hline

\textbf{Shared memory disabled }         &                     &                         \\ \hline
sieve (35,981,314 integers)     & 147.0                 & 828.4                   \\ \hline
HAVi-asyn (77,043,711 integers) & 399.2               & 1839.1                  \\ \hline
ABP (178,172,465 integers)      & 651.9               & 3892.8                  \\ \hline \hline
\textbf{Shared memory enabled }          &                     &                         \\ \hline
sieve (35,981,314 integers)     & 2022.7              & 1271.7                  \\ \hline
HAVi-asyn (77,043,711 integers) & 5131.4              & 2734.0                    \\ \hline
ABP (178,172,465 integers)      & 8251.4              & 6352.4                  \\ \hline
\end{tabular}
\end{table}

In Table \ref{tab:nonrandomres} the results of executing the performance evaluation on the sequences of non-random data can be found. Again a significant difference between having shared memory enabled and disabled exists. Enabling shared memory doubles the running time of most tests, and in some cases the version with shared memory disabled is even ten times faster. \par
As the same state is often encountered during state space exploration all three non-random sequences contain duplicate elements. For Sieve and HAVi-asyn an integer in the sequence occurs on average five times in the sequence. For ABP an integer occurs on average eight times in the sequence. \par
When analysing the performance of the hash table implementations with shared memory disabled, it becomes clear that Cuckoo hashing again has a better performance. However, for non-random data this performance difference appears to be even larger than for random data, as Cuckoo hashing tends to be roughly six times faster than \GPUexplore hashing.

\section{Conclusions and future work}

In this paper we analysed the performance of two lock-free hash table implementations on the GPU. To conduct this analysis we utilised sequences of both random and non-random numbers and measured the time taken to insert all elements in those sequences. \par 
From the performance evaluations in Section~\ref{sec:results} we can conclude that Cuckoo hashing is roughly three times faster than \GPUexplore hashing for random sequences. Apparently the uncoalesced memory accesses performed by the Cuckoo hashing kernel do not impact performance so much that it performs worse than the coalesced memory accesses performed by \GPUexplore hashing. \par 
Even more notable is the performance difference when using non-random data representing sequences of discovered states from actual state space explorations. There are sequences for which Cuckoo hashing is ten times faster than \GPUexplore hashing. This indicates that it is very interesting to investigate whether it is possible to use Cuckoo hashing to further speed up \GPUexplore.
A possible explanation for this performance difference between random and non-random data is that Cuckoo hashing performs better when several duplicate elements occur after each other in the sequence. This is a pattern that is often seen during state space exploration, as a state space generally tends to consist of strongly connected components~\cite{propertiesStateSpaces}. \par 
Furthermore, allocating shared memory when launching kernels appears to have a significant negative impact on the performance of hash table insertions. Therefore, we conclude that for state space exploration, it is worth investigating whether splitting the work of exploring and storing states into different kernels may improve the overall performance of model checking. In such an approach, it would be possible to perform the exploration of states in a kernel with shared memory enabled, allowing for the storage of intermediary results shared memory until no more space is available, and the content of the shared memory is copied to the global memory. Then, a second kernel could be started with shared memory disabled that actually inserts the newly discovered states into the hash table. 

However, even though \GPUexplore hashing uses 32 threads to insert a single integer it is not 32 times slower than Cuckoo hashing. This might be due to the fact that the memory accesses of \GPUexplore hashing are more structured, as \GPUexplore hashing accesses global memory in coalesced blocks of 32 integers. It might be possible to identify a balance between the two approaches, by conducting experiments with \GPUexplore hashing using smaller buckets. With smaller buckets, fewer threads are used to insert a single item, while the advantage of coalesced memory accesses is still preserved. Consequently, in~\cite{gpuexplore-scalability}, we analyse the performance of \GPUexplore hashing with smaller bucket sizes. \par 
Finally, the current performance evaluations have been executed on sequences of 32-bit integers. However, \GPUexplore uses vectors of 32 bit integers as state identifiers, and this is required to perform model checking. Hence, for the purpose of state space exploration it is also interesting to try to modify Cuckoo hashing to support keys of arbitrary length, and evaluate the performance between such a generalised Cuckoo hash table and \GPUexplore hashing.

\nocite{*}
\bibliographystyle{eptcs}
\bibliography{paper}
\end{document}